\documentclass[useAMS,usenatbib,letters]{mn2e}
\usepackage{graphicx}
\def\gsim{\mathrel{\raise.5ex\hbox{$>$}\mkern-14mu
             \lower0.6ex\hbox{$\sim$}}}
\def\lsim{\mathrel{\raise.3ex\hbox{$<$}\mkern-14mu
             \lower0.6ex\hbox{$\sim$}}}

\newcommand{\aap}{    {\it Astron. Astrophys.}}

\newcommand{\apj}{    {\it Astrophys. J.}}
\newcommand{\apjl}{    {\it Astrophys. J. Lett.}}

\newcommand{\jpp}{    {\it J. Plasma Phys.}}
\newcommand{\mnras}{  {\it Mon. Not. Roy. Astron. Soc.}}

\title[NM]{Are there two types of pulsars?}
\author[Contopoulos]{I. Contopoulos$^{1,2}$\\
$^{1}$Research Center for Astronomy and Applied Mathematics,
Academy of Athens, Athens 11527, Greece\\
$^{2}$National Research Nuclear University, 31 Kashirskoe highway, Moscow 115409, Russia}

\begin{document}

\date{Accepted ... Received ...; in original form ...}

\pagerange{\pageref{firstpage}--\pageref{lastpage}} \pubyear{2015}

\maketitle

\label{firstpage}

\begin{abstract}
In order to investigate the importance of dissipation in the pulsar magnetosphere we decided to combine Force-Free with Aristotelian Electrodynamics. We obtain solutions that are ideal (non-dissipative) everywhere except in an equatorial current sheet where Poynting flux from both hemispheres converges and is dissipated into particle acceleration and radiation. We find significant dissipative losses (up to about $50\%$ of the pulsar spindown luminosity), similar to what is found in global PIC simulations in which particles are provided only on the stellar surface. We conclude that there might indeed exist two types of pulsars, strongly dissipative ones with particle injection only from the stellar surface, and ideal (weakly dissipative) ones with particle injection in the outer magnetosphere and in particular at the Y-point.
\end{abstract}


\section{Ideal force-free magnetospheres}

For several decades, people have studied the pulsar magnetosphere using the formalism of Force-Free Electrodynamics (hereafter FFE). It has been assumed that electric charges are `freely' supplied anywhere they are needed in order to satisfy the force-free condition, namely
\begin{equation}
\rho_e {\bf E} + \frac{1}{c}{\bf J}_{\rm FFE}\times {\bf B}=0\ .
\label{FF}
\end{equation}
In doing so, any electric field component parallel to the magnetic field is shorted out and
\begin{equation}
{\bf E}\cdot {\bf B}=0
\label{EperpB}
\end{equation}
everywhere in the magnetosphere. Here, $\rho_e\equiv \nabla\cdot{\bf E}/4\pi$ and ${\bf J}_{\rm FFE}$ are the electric charge and current densities respectively, and $c$ is the speed of light. Eq.~(\ref{FF}) together with Maxwell's equations yields the following expression for the electric current density
\begin{equation}
{\bf J}_{\rm FFE} = \rho_e c\frac{{\bf E}\times {\bf B}}{B^2}+c\frac{{\bf B}\cdot (\nabla\times{\bf B})+{\bf E}\cdot (\nabla\times{\bf E})}{B^2}{\bf B}
\label{JFFE1}
\end{equation}
\citep{Gruzinov2006}. The work of several researchers has shown that the FFE prescription leads to the development of electric current sheets inside which the force-free condition breaks down \citep{1999ApJ...511..351C, 2006ApJ...648L..51S, 2009A&A...496..495K}. There are ways to numerically treat current sheets as contact discontinuities, and, as a result, numerical FFE simulations show little magnetospheric dissipation (less than about $20\%$ of the spindown luminosity) which has mostly been attributed to numerical effects. Still, FFE is not a self-consistent description of the pulsar magnetosphere, and this is why several authors opted for other approaches such as resistive \citep{2012ApJ...754L...1K, 2012ApJ...746...60L}, SFE \citep[Strong Field Electrodynamics;][]{Gruzinov2008}, Aristotelian \citep[particles moving at the speed of light and experiencing radiation reaction opposite to their motion;][]{Gruzinov2013}, `ab initio' global PIC (Particle-In-Cell) simulations, etc. It is interesting to notice that PIC simulations with copious particle injection mimicking pair formation everywhere show little magnetospheric dissipation (less than about $20\%$ of the spindown luminosity), which is consistent with ideal FFE simulations \citep[e.g.][]{2014ApJ...785L..33P, 2016MNRAS.457.2401C}. On the other hand, PIC simulations with particle injection only from the stellar surface show higher amounts of magnetospheric dissipation \citep[$30-50\%$ of the spindown luminosity; e.g.][]{2014ApJ...795L..22C, 2015ApJ...801L..19P}.

We believe that the latter result is physically significant and not a numerical artifact, and in fact, the goal of the present work is precisely to help us understand whether indeed pulsar magnetospheres fall into two categories, weakly and strongly dissipative ones. Before we proceed, we would like to remind the reader how FFE is implemented numerically.  Eq.~(\ref{JFFE1}) can be rewritten as
\begin{equation}
{\bf J}_{\rm FFE} = \rho_e c\frac{{\bf E}\times {\bf B}}{B^2}+\frac{\partial}{\partial t}({\bf E}\cdot{\bf B})\frac{\bf B}{B^2}+({\bf J}\cdot{\bf B})\frac{\bf B}{B}\ .
\label{JFFE}
\end{equation}
The middle term in the r.h.s. of Eq.~(\ref{JFFE}) obviously disappears if Eq.~(\ref{EperpB}) is implemented. The third term is the part of the electric current that guarantees satisfaction of Eq.~(\ref{EperpB}). In practice, we implement the FFE prescription as follows:
\begin{enumerate}
\item We set
\begin{equation}
{\bf J}_{\rm num}= \rho_e c\frac{{\bf E}\times {\bf B}}{B^2}\ .
\label{Jnum1}
\end{equation}
Notice that the electric current component parallel to ${\bf B}$ is missing.
\item We use Maxwell's equations to advance ${\bf B}$ and ${\bf E}$ to the next time step using the above expression for the electric current.
\item We correct ${\bf E}$ by removing from it any nonzero component parallel to ${\bf B}$ that accumulated during the above step.
\end{enumerate}
Therefore, the implementation of Eq.~(\ref{EperpB}) is a central procedure in all FFE numerical codes. What is not so obvious in the FFE formalism is that we must add one more constraint, namely that
\begin{equation}
\ \ \ \ \mbox{(iv)}\ \ E\leq B\ .
\label{ElB}
\end{equation}
The latter does not come about naturally from the FFE equations and must be imposed to avoid superluminal drift velocities.

As is by now well known, when we integrate the FFE equations, current sheets appear in the magnetosphere. 
These regions are numerically very problematic and every code addresses them in its own way. The extra constraint of Eq.~(\ref{ElB}) is crucial. Our personal experience has shown that during the numerical integration, magnetic field lines are stretched open and a region appears in the magnetosphere where the magnetic field drops to zero. If we enforce Eq.~(\ref{ElB}), a contact discontinuity with $B=0$ inside it (i.e. with full magnetic field reversal accross it) grows beyond that region. Otherwise, a different type of contact discontinuity forms with $B\neq 0$ inside it (i.e. with magnetic field threading it; see next section).

Also, step (iii) above can only be implemented iteratively throughout the numerical grid. Our personal experience has shown that this is numerically unstable, and the programmer must find the optimal number and frequency of iterations to suppress the instability. This introduces spurious numerical dissipation in the equatorial current sheet, and one must test the numerical code by comparing its results with the few known high-resolution axisymmetric solutions where indeed current sheets appear as mathematical contact discontinuities \citep[see for example][for a discussion of problems with the FFE treatment]{2006MNRAS.367...19K}.

Current sheets are thus treated differently in different codes, and this may be one reason why several researchers opted for global PIC simulations that resolve current sheets.
We on the contrary believe that it is too early for `ab initio' magnetospheric simulations \citep[for an extended presentation of our arguments see][]{Contopoulos2016}, and we decided instead to investigate the potential of more general formalisms that accomodate current sheets self-consistently. In particular, there is no reason to require that $E\leq B$ inside a current sheet. We thus need to consider a formulation that allows for $E$ to surpass $B$. One promising such alternative may be derived from Aristotelian electrodynamics. 

\section{Dissipative magnetospheres}

Aristotelian electrodynamics \citep[hereafter AE;][]{1989A&A...225..479F} assumes that the charged positive and negative particles that fill the magnetosphere move at (in fact almost at) the speed of light with velocities
\begin{equation}
{\bf v}^\pm = c\frac{{\bf E}\times {\bf B}\pm (E_0{\bf E}+B_0{\bf B})}{B^2+E_0^2}\ ,
\label{vAE}
\end{equation}
and that force-balance must include the radiation reaction force along and opposite to the direction of the particle velocity. Here, $B_0$ and $E_0$ are the magnitudes of the magnetic and electric field in the frame in which the two are parallel to each other and the particles move along them at the speed of light. The expression for the electric current density becomes
\begin{equation}
{\bf J}_{\rm AE} = \frac{\rho_e c{\bf E}\times {\bf B}+\rho_0c(E_0{\bf E}+B_0{\bf B})}{B^2+E_0^2}\ ,
\label{JAE}
\end{equation}
where, $\rho_0\equiv \rho_+ + |\rho_-|$ is the sum of the electric charge densities of the two types of magnetospheric particles. Eq.~(\ref{JAE}) is not by itself sufficient to solve for the electromagnetic field structure, and one needs to know the sources of the particles that populate the electric current, thus making the problem complicated and not fully constrained. In particular, Andrei Gruzinov who in the past four years has been the main proponent of AE was the first to discuss the significance of the plasma production process in determining the global structure of the magnetosphere \citep{Gruzinov2013}. He argued that, when particles are supplied only on the stellar surface, he obtains solutions with large amounts of dissipation ($30-50\%$ of the spindown luminosity). On the contrary, when particles are freely supplied anywhere in the magnetosphere they are needed \citep[as in][]{2014ApJ...785L..33P, 2016MNRAS.457.2401C}, the solutions are closer to ideal FFE with small amounts of dissipation (less than about $20\%$ of the spindown luminosity). 

We are determined to understand the main difference between the above two classes of solutions. In order to show how AE tranforms Poynting flux into particle energy, Gruzinov offered a simple example that does not require the specification of the rate of particle production, his so-called `Device' \citep{Gruzinov2014}. 
This example contains only one type of charge particles, is exactly solvable, and was specifically chosen to represent the annihilation of colliding Poynting fluxes in a non-force-free radiation zone bounded by two force-free zones, in analogy to real pulsars.
It is interesting that in his example, ${\bf E}$ is everywhere perpendicular to ${\bf B}$. In particular, ${\bf E}\cdot {\bf B}=0$ not only in the FFE zone, but also in the AE dissipative radiation zone. This is indeed a simple generalization of the FFE current sheet contact discontinuity. We thus decided to follow his lead and implement a prescription where ${\bf E}\cdot {\bf B}=0$ {\em everywhere} in the magnetosphere. In that special case, the prescription of Eq.~(\ref{JAE}) simplifies considerably since $E_0=0$ if $E\le B$, and $B_0=0$ if $E>B$. We have generalized the FFE numerical prescription as follows:
\begin{enumerate}
\item We set
\begin{equation}
{\bf J}_{\rm num}= \frac{\rho_e c{\bf E}\times {\bf B}+|\rho_e |cE_0{\bf E}}{B^2+E_0^2}\ .
\label{Jnum2}
\end{equation}
Here, $E_0$ is either zero if $E\le B$, or is equal to $(E^2-B^2)^{1/2}$ if $E>B$. Notice that the electric current component parallel to ${\bf B}$ is missing, as was the case in FFE too.
\item We use Maxwell's equations to advance ${\bf B}$ and ${\bf E}$ to the next time step using the above expression for the electric current.
\item We correct ${\bf E}$ by removing from it any nonzero ${\bf E}\parallel {\bf B}$ component that accumulated during the above step.
\end{enumerate}
The implementation of Eq.~(\ref{EperpB}) is thus a central procedure that either yields FFE in regions where $E\le B$, or transitions to AE with only one type of charge in regions where $E>B$. The numerical integration determines self-consistently where the AE dissipative radiation zones will develop. As we will see next, the new solutions are qualitatively similar to our `New Standard Magnetosphere' \citep{2014ApJ...781...46C}, and also to the results of much more complex large scale PIC simulations with pair supply only from the stellar surface.

\begin{figure*}
\centering
\vspace{2.5cm}
\begin{minipage}[b]{0.5\textwidth}
\centering
\includegraphics[trim={2.5cm 2cm 4.5cm 6cm}, width=0.4\textwidth]{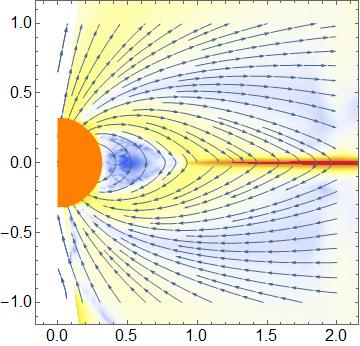}
\end{minipage}%
\hspace{-0.5cm}
\begin{minipage}[b]{0.5\textwidth}
\centering
\includegraphics[trim={2.5cm 2cm 4.5cm 6cm}, width=0.4\textwidth]{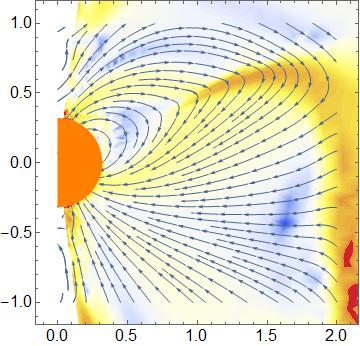}
\end{minipage}%
\vspace{5.7cm}
\begin{minipage}[b]{0.5\textwidth}
\centering
\includegraphics[trim={2.5cm 2cm 4.5cm 6cm}, width=0.4\textwidth]{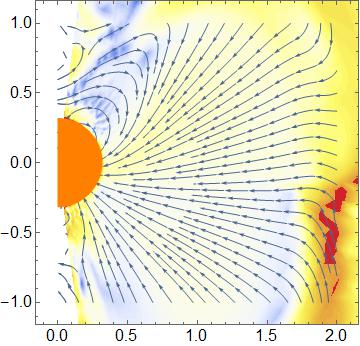}
\end{minipage}%
\hspace{-0.5cm}
\begin{minipage}[b]{0.5\textwidth}
\centering
\includegraphics[trim={2.5cm 2cm 4.5cm 6cm}, width=0.4\textwidth]{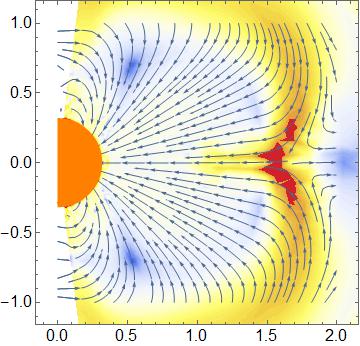}
\end{minipage}%
\vspace{1.5cm}
\caption{Results of 3D numerical simulations of dissipative magnetospheres after 1.75 rotations for various pulsar inclinations (top left: $0^\circ$, top right: $30^\circ$, bottom left: $60^\circ$, bottom right: $90^\circ$). Distances in units of the light cylinder radius $r_{\rm LC}$. Lines with arrows: magnetic field lines on the poloidal (meridional) plane. Color scale: the magnitude of the poloidal component of $r^2\nabla\times {\bf B}$ (logarithmic scale; blue: $\sim$~zero; red: $\sim$~50 times $B$ at the light cylinder multiplied by $r_{\rm LC}$). The current sheet (red and dark orange) originates on the Y-point and extends outwards.}
\label{Fig1}
\end{figure*}

\subsection{Solutions}

We integrate the time-dependent Maxwell's equations with our new prescription for the electric current. Our numerical code is written in $(r, \theta, \phi)$ spherical coordinates with $1^\circ$ resolution in $\theta$,  $2^\circ$ resolution in  $\phi$, $2:1$ grid cell aspect ratio in the $r$ and $\theta$ directions, and a $5^\circ$ avoidance grid zone around the rotation axis $\theta=0^\circ$). We acknowledge that these are our first low resolution crude simulations only meant to support our point about the importance of dissipation in the pulsar magnetosphere. In Fig.~1 we show the magnetic field and electric current structures on the poloidal plane for various pulsar inclination angles after three full rotations. In all cases, the numerical integration started with a magnetostatic dipole, and then the central `star' was set to uniform rotation. The most important thing to remember is that the new prescription {\em does not} restrict the magnitude of ${\bf E}$ to be less than that of ${\bf B}$. It only requires that the two are everywhere perpendicular to each other.

It is interesting that our force-free region extends all the way from the axis to the current sheet, i.e. we do not obtain the transition from FFE to AE at intermediate latitudes found by Gruzinov who implemented the same AE formulation everywhere. What we are doing is different since we have effectively relaxed the requirement that plasma particles move at the speed of light in force-free regions (Ted Jacobson, private communication).

All solutions shown are strongly dissipative in the equatorial region around the current sheet beyond the light cylinder. In that respect, our solutions are qualitatively similar to the FIDO solutions (FFE inside, dissipative outside) of \citet{2014ApJ...793...97K}. We are still unable to reliably determine the total amount of dissipation since our present numerical simulations do not extend far enough beyond the light cylinder. We estimate that, in the aligned rotator roughly $30\%$ of the total spindown luminosity is dissipated between one and two light cylinder radii, and $50\%$ up to four light cylinder radii.\footnote{Notice that without the second term in the numerator of Eq.~(\ref{Jnum2}), the solutions are slightly less dissipative ($20\%$ of the  spindown luminosity is dissipated between one and two light cylinder radii), and contain a separatrix current sheet as in FFE solutions (see below).}

The most important result of our present work is that the return current is spread smoothly, and does not reach the star along one particular surface (in FFE this coincided with the boundary of the corotating `dead' zone, the so-called separatrix). A current sheet develops {\em only} in the equatorial region, and its electric current consists of the accumulated electric current that flows along the FFE field lines that enter it. This configuration is qualitatively very similar to our `New Standard Magnetosphere' \citep{2014ApJ...781...46C}, in which we had required that the electric current of the equatorial current sheet does not cross the Y-point and continue along the separatrix. We define the Y-point (Y-line in 3D) as the tip of the so called `dead' zone (the corrotating closed line region in which no Poynting flux flows towards the outer magnetosphere). Our main argument in that paper has been that at the Y-point the electric charge density drops to zero, thus if an electric current crossed the Y-point, the current sheet would not be supported electrostatically. Furthermore, in \citet{Contopoulos2016} we argued that it is very hard for particles of the charge type required to support the separatrix current sheet (electrons in the case of aligned pulsars, positrons/protons/ions in the case of counter aligned ones) to cross the Y-point from outside the light cylinder. Our reasoning has been twofold. Firstly, their Speiser orbits are `messy' because when the particles reach the upper and lower outskirts of the current sheet, the overall ${\bf E}\times {\bf B}$ direction is outwards and not inwards. Secondly, in 3D, the equatorial current sheet outside the light cylinder has an undulating corrotating trailing spiral shape along which particles can only move outwards, not inwards. The above theoretical conclusions seem to be in agreement with the results of global PIC simulations with pair supply only from the stellar surface \citep{2015MNRAS.448..606C, 2015ApJ...801L..19P}.

Our above arguments for the vanishing of the separatrix electric current sheet {\em break down} if a) copious pair production at the Y-point supplies the necessary charge carriers to support both the equatorial and separatrix current sheets \citep[as is the case for example in][]{2014ApJ...785L..33P, 2016MNRAS.457.2401C}, and b) if a thermal particle population supports the current sheet at the Y-point \citep[see for example][]{2014ApJ...780....3U}. The latter possibility seems less likely for relativistic leptons which have radiated away their random motion component. This leads us to agree with \citet{Gruzinov2013} 
who suggested that there are two types of pulsars, `strong' ones with copious pair production in the outer magnetosphere and in particular at the Y-point, and `weak' ones with particle supply only from the stellar surface. In the latter case, the supplied particles may very well be electrons and protons or ions, which would make aligned weak pulsars very different from counter-aligned weak ones. The latter theoretical conclusion remains to be tested observationally. Notice that, in Gruzinov's nomenclature, `weak' pulsars are strongly dissipative, and `strong' pulsars weakly!

\section{Summary}

We have found out that if we combine FFE with a particular limit of AE, namely one containing only one type of charge carriers, we obtain a dissipative equatorial current sheet that develops beyond the light cylinder in an otherwise ideal pulsar magnetosphere. Poynting flux converges into that region from above and below and is transformed into particle acceleration and radiation, as in Gruzinov's `Device'. The current sheet originates at the Y-point and does not extend along the separatrix to the stellar surface. Its electric current consists of the accumulated charged particles that flow along the field lines that enter it (obviously, only one type of charge is being supplied from the stellar surface along these field lines). Our solutions are qualitatively similar to our `New Standard Magnetosphere' and may be representative of so-called `weak' pulsars in which particles are supplied only at the stellar surface and not in the outer magnetosphere. They are strongly dissipative, thus different from the standard ideal FFE weakly dissipative solutions representative of so-called `strong' pulsars in which copious pair production must take place in the outer magnetosphere and in particular at the Y-point. Intermittent pulsars offer a hint that indeed more than one pulsar configuration may be realized in nature.

Our present numerical simulations do not extend far enough beyond the light cylinder for us to obtain reliable estimates of the amount and distribution of magnetospheric dissipation for various pulsar inclination angles.
There is preliminary evidence that the dissipative current sheet may become unstable and not survive beyond a few light cylinder radii, but we need to investigate this issue further. On the other hand, we are not in a position to estimate how much our (and others') particular numerical treatment of the current sheet suppresses dynamic magnetospheric features such as current sheet instabilities, ongoing reconnection, plasmoid formation and detachement, etc. We plan to obtain more detailed results in a future publication. We believe that we need to invest more effort, both theoretical and numerical, in the investigation of the detailed magnetospheric structure and physical conditions around the Y-point which may hold the key to the determination of the global amount of magnetospheric dissipation.


\section*{Acknowledgements}
We would like to thank the hosts and participants of the $2^{\rm nd}$ Purdue Workshop on Relativistic Plasma Astrophysics, in particular Drs. Maxim Lyutikov, Andrei Gruzinov, Jonathan Arons, Anatoly Spitkovsky and Alex Chen, who helped us challenge the standard picture of an ideal dissipationless pulsar magnetosphere. 
We would also like to thank Dr. Ted Jacobson who pointed out to us that what we are using a limited form of Aristotelian Electrodynamics. Our numerical simulations were performed using the resources of the NRNU MEPhI High-Performance Computing Center.

\bibliographystyle{mnras}
\bibliography{NewMagnetosphere}

\end{document}